\documentclass{aa}
\usepackage{graphics}
\usepackage{times}                   

\long\def\jumpover#1{{}}

\def \approxgt{\,\raise2pt \hbox{$>$}\kern-8pt\lower2.pt\hbox{$\sim$}\,}
\def \approxlt{\,\raise2pt \hbox{$<$}\kern-8pt\lower2.pt\hbox{$\sim$}\,}
\def \th{\thinspace}
\def \ngth{\negthinspace}
\def \ngth2{\negthinspace\negthinspace}
\def \moi{\negthinspace -\negthinspace}

\def \ni{\noindent}
\def \Teff{{$T_{\rm {ef\!f}} $}}
\def \teff{{T_{\rm {ef\!f}} }}

\def\Lo{{$L_\odot $}}

\def \eg{{{\it e.g.},\ }}
\def \etal{{\it et al.}}

\def \cf{{\it cf.\ }}

\def \ie{{{\it i.e.},\ }}

\def \viz{{\it viz.\ }}
\def \vs{{\it vs.\ }}
\def\Log{{\mathrm {Log}}}

\def \dotm{\hbox{$.\!\!^{\rm m}$}}

\begin{document}



\title{Cepheid mass-luminosity relations from the Magellanic Clouds}
\titlerunning{Mass-luminosity relations from the Magellanic Clouds}
\authorrunning{Beaulieu \etal}
\author{J. P. Beaulieu\inst{1},
J.R. Buchler\inst{1,2},  
Z. Koll\'ath\inst{1,3} }
\institute{
Institut d'Astrophysique de Paris, 98bis Boulevard Arago, 75014 Paris, FRANCE
\and Department of Physics, University of Florida, Gainesville, Fl 32611, USA
\and
Konkoly Observatory, POB 67, Budapest, 1525 HUNGARY}

\abstract{
The OGLE data base is used in conjunction with Kurucz atmosphere
models to generate sets of period, effective temperature and
luminosity for fundamental and overtone Magellanic Cloud Cepheids.
The Florida pulsation code (with linear turbulent convection) is
then used to compute masses for these stars, assuming an average
composition of ($X$=0.716, $Z$=0.010) for the LMC and of ($X$=0.726,
$Z$=0.004) for the SMC.  The average $M$--$L$ relation for the
fundamental Cepheids matches closely that for the first overtone
Cepheids for each Magellanic Cloud.  Neither the SMC nor the
LMC average $\Log M$--$\Log L$ relations are straight, but have a
noticeable curvature.
In view of the uncertainties in distance and reddening we have
adopted three different choices for these quantities. The results
based on the 'long' distance scale to the clouds give a better
agreement between theory and and observations than the 'short' one.
All the current evolutionary tracks predict systematically larger
masses for given luminosities than our observationally derived ones,
especially at the high end.  Moreover, our study confirms that the
evolutionary tracks of the low mass stars in SMC are not in agreement with
the observations as they do not extend sufficiently blueward and do
not penetrate deep enough into the instability strip, or not at all.
The inference of masses directly from the observational database
yields a novel and strong constraint on evolutionary calculations.
\keywords {stars : oscillations -- stars: Cepheids 
 -- Stars: Evolution, Magellanic Clouds, distance moduli}}
\maketitle

\section{Introduction}

In the last few years, high quality data on large numbers of Cepheid
variables in the Small and Large Magellanic Clouds have been made
available by the EROS and OGLE microlensing projects (Beaulieu
\etal\ 1995, Afonso \etal\ 1999, Udalksi \etal\ 1999abc).  In particular the
OGLE Project has provided standard colors in addition to periods and
magnitudes for the largest samples published to date.  In this paper
we examine some of the constraints that the MC Cepheids impose on
stellar evolution and stellar pulsation theories.

 \begin{figure*}
 \centerline{\resizebox{9.7cm}{!}{\includegraphics{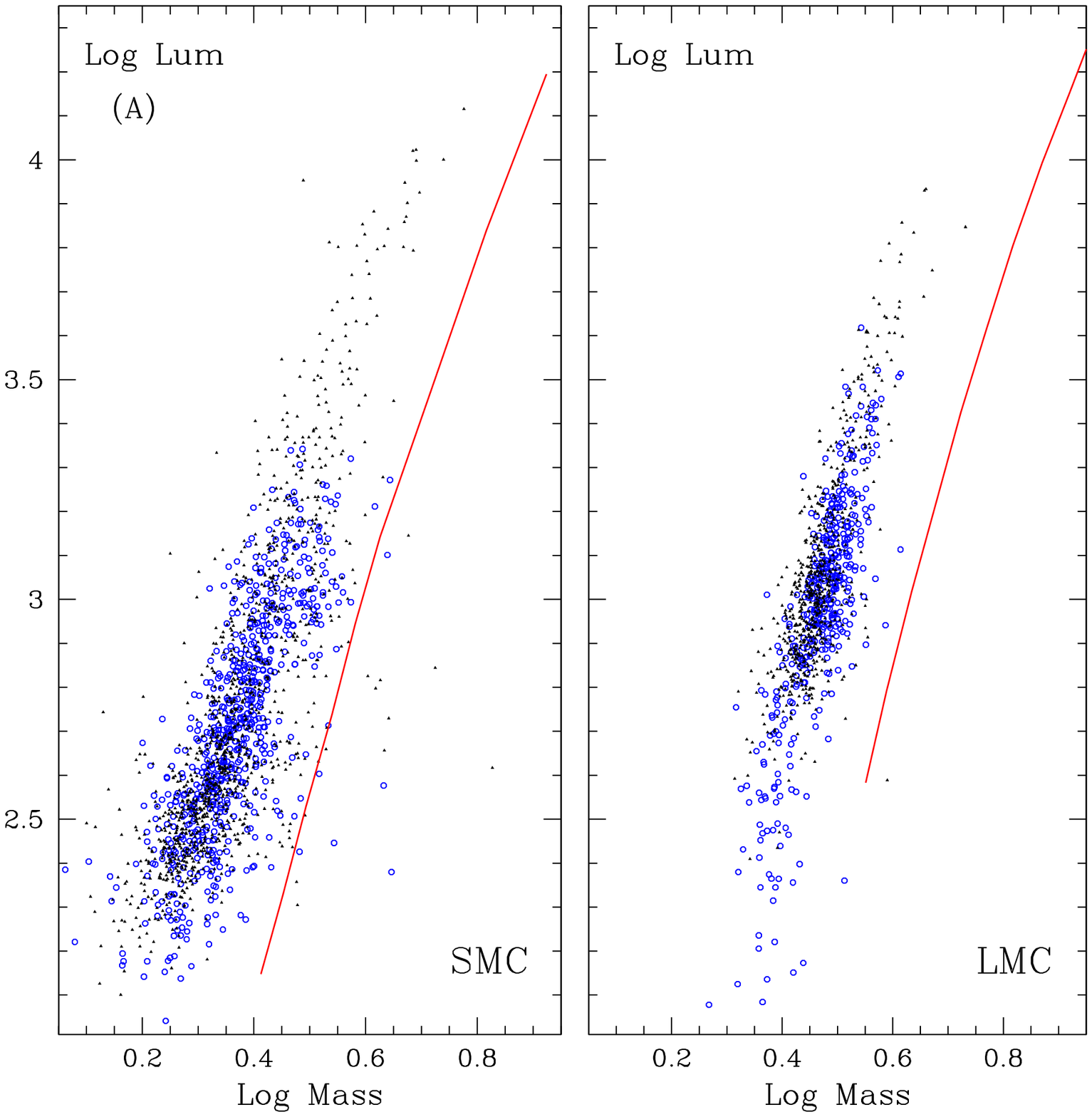}}\hskip -20pt
\resizebox{9.7cm}{!}{\includegraphics{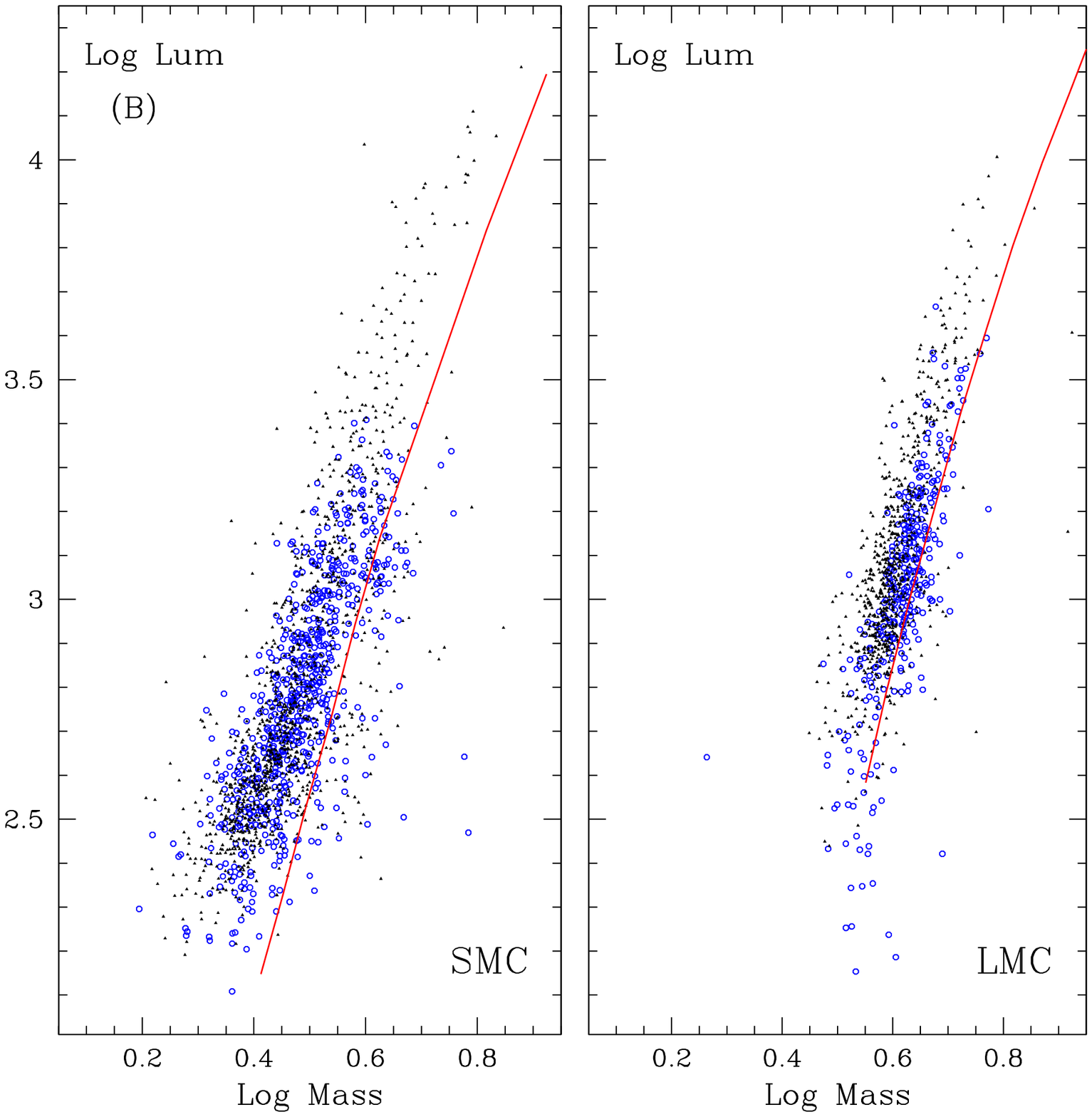}}}
 \hfill
 \vskip -10pt
 \parbox[b]{17cm}{
\caption[]{\small Mass--Luminosity relations for SMC  and LMC as
 derived from the OGLE data adopting choice 
(A) for distance modulus and reddenings in the left panel 
and our preferred choice (B) in the right panel;
 fundamental Cepheids are shown as dots and overtones as open circles. 
As a guide to the eye we have 
superposed the evolutionary $M$-$L$ relations of Girardi \etal.
   }
 \label{mcab}
 }
 \end{figure*}

We use the full catalogue of publicly available of LMC and SMC
single mode Cepheids and SMC double mode Cepheids produced by OGLE
in BVI (Udalski \etal, 1999abc, with the zero point corrections as
suggested in the April 2000 OGLE web site, U99 hereafter).  The
single mode Cepheid catalogues contain 1435 LMC and 2167 SMC stars.
We keep the objects classified as fundamental mode pulsators or
first overtone pulsators, with reliable photometry in both V and I.
We exclude stars whose magnitudes are most likely to be strongly
contaminated by companions or blending in V or I.  The remaining
stars form our working sample of OGLE Cepheids.  It consists of 670
LMC fundamentals, 426 LMC overtones, 1197 SMC fundamentals and 677
SMC overtones, as well as 24 F/O1 SMC double-modes and 71 O1/O2 SMC
double-modes.

The OGLE data base provides intensity averaged magnitudes and
colors.  With the help of distance moduli, these can then be
transformed to luminosities and effective temperatures.  The
Magellanic Clouds (MC) are thought to be relatively uniform in
composition, and with the observed average compositional
information theoretical modeling can then provide the mass of each
star.

 \begin{figure}
 \centerline{\resizebox{9.7cm}{!}{\includegraphics{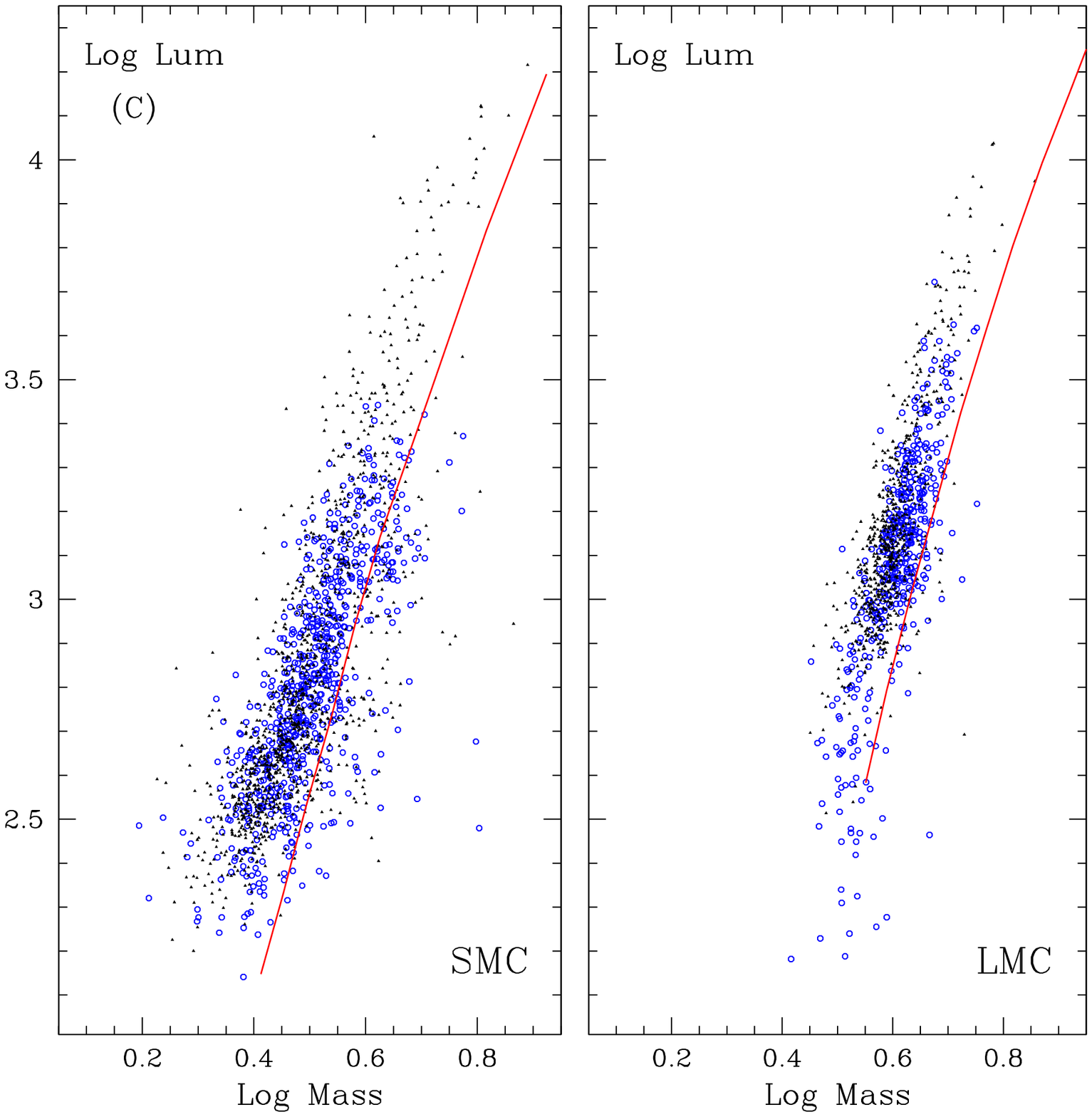}}}
 \hfill
 \vskip -10pt
 \parbox[b]{8.cm}{
\caption[]{\small Mass--Luminosity relations for SMC  and LMC as
 derived from the OGLE data with choice 
(C) for distance modulus and reddenings.   
 Superposed are the $M$-$L$ relations of Girardi \etal.
   }
 \label{mcc}
 }
 \end{figure}

We have to adopt distances and reddenings for the Magellanic
Cepheids.  The distance of the Large Magellanic Cloud remains at the
center of the current debate about the distance scale ladder.
Extreme values of the distance modulus range from 18.08 to 18.70,
but the distance estimates tend to cluster around 18.3 for the
'short' distance scale, and around 18.5 for the 'long' distance
scale (see Walker 1999, Udalski 2000 and references therein, Cole
1998, Girardi 1998, Stanek \etal\ 2000, Romaniello \etal\ 2000, Feast
\& Catchpole 1997, Luri \etal\ 1998, Groenewegen \& Oudmaijer, 2000
and references therein, Groenewegen 2000, Sakai \etal\ 2000, Cioni \etal\ 2000).  It is
beyond the scope of the paper to solve this distance scale problem,
but we will justify the choices we adopt for distances and
reddenings.

U99 derive reddening by making the assumption that the red clump I
luminosity has a weak metallicity dependence, and they use the I
luminosity to map relative variations of E(B-V).  Finally they fix
the zero point of the relative scales on the basis of observations
of NGC1850, NGC1835, HV2274 in the LMC, and NGC416 and NGC330 in the
SMC.  These assumptions lead to the determination of a 'short'
distance to the LMC of $18.24 \pm 0.05$ and $18.75 \pm 0.05$ to the
SMC.  The relative distance between LMC and SMC obtained with four
different distance indicators (Cepheids, RR~Lyrae, red clump and tip
of the red giant branch) but with the same reddening maps, gives a
relative distance modulus of $0.50 \pm 0.02$. 

The reddening maps of U99 give a mean reddening for the LMC of
0.147, and of 0.092 for the SMC.  The consistency of the zero point
of the reddening scale between these different observations is at
the level of few times $10^{-2}$ (\eg $E(B-V)=0.15 \pm 0.05$ for
NGC1850 and $E(B-V)=0.13 \pm 0.03$ for NGC1835).  We note that these
values are different from what is usually given as mean properties
for the clouds (especially for the LMC, see Walker 1999 and
references therein).  In his recent review, Walker (1999) noted that
the median reddenings are $E(B - V) \sim 0.1$ for the LMC and $E(B -
V) \sim 0.08$ for the SMC.  The galactic foreground reddenings along
the line of sight of the clouds are known to be low, \viz 0.06 and
0.04, respectively. The estimation of differential reddening inside
the clouds based on earlier studies is quite uncertain.  In
particular, heavily reddened stars ($E(B- V)=0.30$) can be found all
over the LMC, but the typical range is 0 -- 0.15.

In view of these uncertainties and controversies, we consider three
alternative choices for the distance moduli and for the reddening
corrections for our derivation of stellar parameters from the OGLE
data: \\
\ni Choice (A) adopts both the distance moduli and the reddening
as suggested by U99. \\
\ni Choice (B) adopts the Cepheid distance
modulus to the LMC of $18.55\pm0.10$, and a relative distance
between LMC and SMC of $0.42\pm 0.05$ (Laney \& Stobie, 1994), with
the mean reddenings of $E(B- V)=0.1$ and $E(B- V)=0.08$ for LMC
and SMC respectively. \\
\ni Choice (C) is intermediate in that it adopts U99's relative 
distance between LMC and SMC 
of $0.50 \pm 0.02$ and its reddening, but a 'long' LMC distance modulus of 18.55.

To summarize the differences, (B) -- (A), in distance moduli 
and in mean reddenings are
$\delta \mu  = 0\dotm 31$,
$\delta \langle E(B - V) \rangle \sim 0.047$ for the LMC, and 
$\delta \mu  = 0\dotm 22$,
$\delta \langle E(B - V) \rangle \sim 0.012$ for the SMC.
The differences, (C) -- (A), in distance moduli and in mean reddenings are
$\delta \mu  = 0\dotm 25$,
$\delta \langle E(B - V) \rangle = 0$ for the LMC and the SMC.

We follow Kov\'acs (2000) in the conversion from magnitudes to bolometric, and
from colors to effective temperatures.  Because of the wider range of masses
that are needed for this study, we have redone the Kov\'acs fits with the
Kurucz (1995) stellar atmospheres from the BaSeL database (Lejeune et al.,
1997) for the LMC and SMC chemical compositions for the temperature, luminosity
and $\log g$ range appropriate for Cepheids ranging from 100\Lo\ to 20,000\Lo,
with the results

 \begin{eqnarray}
  2.5\th \Log\th L &=& \mu_{MC}- V +R_V\th E(B\moi V) + BC +4.75 \\
  \Log \th g &=& 2.62 - 1.21\th \Log \th P_0 \\
 {\rm SMC:\hfill} \nonumber \\
  \Log \th\teff &=& 3.91611   + 0.0055\th \Log \th g \nonumber \\
       &&\quad - 0.2482 (V\moi I_c \moi (R_V-R_I) E(B\moi V))  \\
  BC &=& -0.0324 + 2.01 \Delta T - 0.0217\th \Log \th g \nonumber\\
             &&\quad - 10.31\th \Delta T^2  \\
 {\rm LMC:\hfill} \nonumber \\
    \Log \th\teff &=& 3.91545   + 0.0056\th \Log \th g \nonumber \\
       &&\quad - 0.2487 (V\moi I_c \moi (R_V-R_I) E(B\moi V))  \\
  BC &=& -0.0153 + 2.122 \Delta T - 0.0200\th \Log \th g  \nonumber\\
             &&\quad - 11.65\th \Delta T^2
 \label{eqs}
 \end{eqnarray}

\noindent where $\Delta T = \log \teff - 3.772$ and $L$ is in solar
units.  The transformation to absolute luminosities is then made
with the adopted distance moduli $\mu_{MC}$ to the LMC or the SMC
for the various choices (A), (B) and (C).  We note that compared to
Kov\'acs there are systematic shifts of 50K in \Teff, and 0.02 in
log $L$. 

It may be objected that since $\log g = \log (G M ) /R^2$ in Eq.~(2)
there is an implicit assumption about a mass--radius --period
relation.  However, in Eqs.~(3 -- 6), $\log g$ appears with a tiny
multiplier, and over the period range from 1 to 10 days its
contributions to Log L and Log \Teff\ vary by 0.013 and 0.0046,
respectively.  Ultimately, there is essentially no feedback on our
mass determination.

\begin{figure*}
\centerline{\resizebox{15cm}{!}{\includegraphics{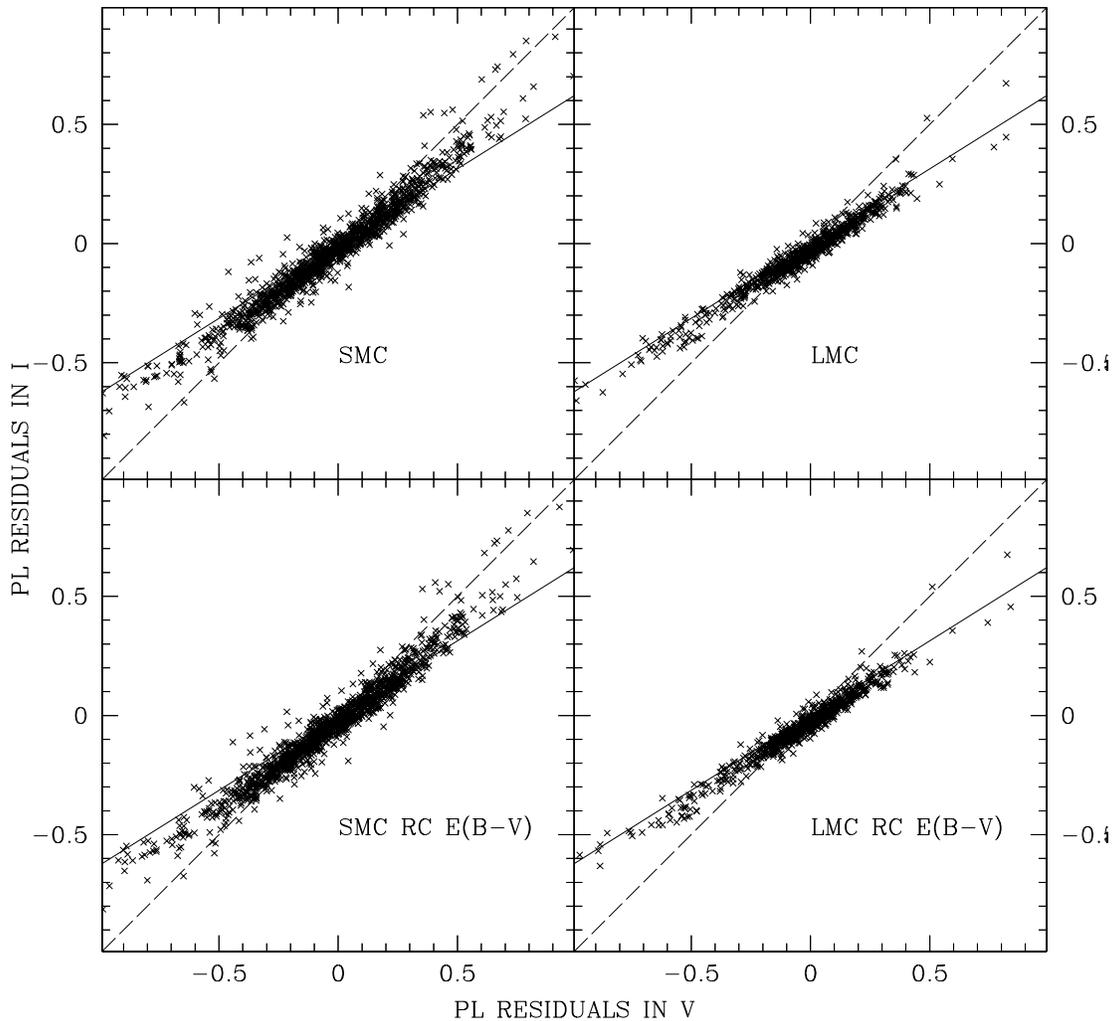}}}
\hfill
\vskip -10pt
\parbox[b]{17cm}{
\caption[]{The two upper panels gives the residuals of the V and I
$P$--$L$ relation for uncorrected reddening in LMC and SMC. The two
lower are giving the residuals of the V and I $P$--$L$ relation
after reddening corrections based on the red clump stars following
the U99 procedure. On each panel are indicated in solid the
reddening line, and in dash the depth dispersion (the diagonal)
line.
}

 \label{res}
 }
\end{figure*}

\section{Mass Luminosity Relation}

With the help of relations (\ref{eqs}) and the OGLE data we
therefore obtain a period, a luminosity and a \Teff\ for each
fundamental and for each overtone Cepheid.  These quantities are
sufficient to compute the masses with the help of a stellar model
builder and linear pulsation code, since $P_0$=$P_0(L,$ $M,$ $
\teff,$ $ X,$ $ Z)$ and $P_1$=$P_1(L, M, \teff, X, Z)$.

For the computation of the SMC and LMC Cepheid models we have adopted the
respective compositions X=0.726, Z=0.004 and X=0.716, Z=0.01.
We have used OPAL opacities (Iglesias and Rogers 1996)
merged with the low temperature ones of Alexander and Ferguson
(1994).  Turbulent convection has been treated as described in
Yecko, Koll\'ath \& Buchler (1998).  The convective parameters were
chosen as in Koll\'ath, Buchler, Szab\'o \& Csubry (2001), although,
the precise values of these parameters will have very
little effect on the periods.

In Fig.~\ref{mcab} we present the $M$--$L$ diagrams obtained from our
Cepheid model calculations that use the observational constraints, A
in the left panel, and B in the right panel.  Fig.~\ref{mcc}
displays the results for choice (C).  The fundamental Cepheids are
represented as dots and the overtone Cepheids as open circles.  To guide
the eye and for later reference we have also shown the $M$--$L$
relations of Girardi \etal\ (2000) for the Cepheids on the second
crossing of the instability strip.

Four features stand out immediately.  First, the three choices give
very similar $M$--$L$ distributions, especially in slope and
scatter, but choice (A) has a zero point that is in substantial
disagreement with the evolutionary calculations.  Second, the
observations indicate a curved mass-luminosity relation.  Third, the
average $M$--$L$ for the fundamental Cepheids agrees with that of
the overtones.  Fourth, there is a huge scatter whose nature needs
to be addressed, because the Cepheids form a homogeneous group,
and one would expect all of them to fall on a very tight $M$--$L$ line.

\section{Discussion}

\subsection{Which choice for distances and reddening ?}

We recall that the main difference between (B) and (C) is the adoption of mean
SMC and LMC reddenings in (B) and (OGLE) reddening maps in (C).  As we have
seen in Figs.~(\ref{mcab}, \ref{mcc} and \ref{res}) there is no significant
improvement with the use of reddening maps, neither in the $P$--$L$ residuals,
nor in the scatter in the final $M$--$L$ plots.  Choices (B) and (C) are
essentially equivalent except for a tiny systematic shift parallel to the
distribution. 

As seen in Fig.~\ref{mcab}, with the observational constraints (A)
there is a systematic shift to higher luminosities and higher
temperatures for the observed stars.  Taken at face value, choice
(A) would indicate that evolutionary calculations are quite off the
beat which we deem unlikely given the general agreement of several
independent calculations (\cf \S4).  The sensitivity of the $M$--$L$
relations to metallicity would have to have been largely
underestimated, too. On the other hand, with the 'long' distances to
the clouds of choices (B) and (C), the situation is much more
satisfactory, as we shall see.

 In the following, unless otherwise specified, we will adopt
choice (B).

\subsection{Computational Uncertainties}

First we examine the computational uncertainties.  We expect these to be small
because we compute only the linear periods of the fundamental and first
overtone.  In contrast to the linear growth-rates, the periods are very
insensitive to the convective parameters ($\alpha$'s in Yecko \etal\ 1998).
The comparison of purely radiative models with our turbulent convective ones
gives an idea of the uncertainty.  We find that the period shifts are
systematic but small, of the order of the size of the dots in Fig.~\ref{mcab}
and ~\ref{mcc}.  The fact that they are systematic indicates that they cannot
contribute to the scatter of Fig.~\ref{mcab} and ~\ref{mcc}.  The models have
been computed with a mesh of 200 points.  Models run with a cruder mesh
distribution give essentially the same $M$--$L$ picture.  We can safely use
linear periods, because nonlinear hydrodynamic modeling shows that the
differences are systematic and at most of the order of 0.1\% which has no
appreciable effect on the $M$--$L$ picture.

We have not been able to find a computational uncertainty that can
account for the scatter in the $M$--$L$ relation, and we have to
look in the observational data.

\subsection{Scatter in the $M$--$L$ Relations}

In the upper panel of Fig.~\ref{res} we plot the residuals of the
period-luminosity ($P$--$L$) relation in V and I for both the LMC
and the SMC fundamental Cepheids OGLE data.  These diagrams
illustrate the structure of the Cepheid P--L relation (see figs. 5
and 6 from Sasselov \etal\ 1997).  The dispersion is mainly along the
reddening vector in the LMC, whereas in the SMC it is not because
depth effects are another source of scatter. We recall (\eg Sasselov
\etal\ 1997) that there is an unfortunate near degeneracy between
lines of constant period and reddening. Therefore one cannot just
minimize the dispersion along the reddening vector in this plane to
correct for the reddening.  It would lead to an overcorrection. When
we use the reddening derived by U99 we note a very marginal
improvement of the residuals as shown in the lower panel of
Fig.~\ref{res} and as noted by U99.

 We conclude that the differential reddening within the clouds on
a star by star basis persists as a major source of dispersion that
is not compensated for by the reddening maps from red clump stars.

\begin{figure}
 \centerline{\hskip 5cm\resizebox{10cm}{!}{\includegraphics{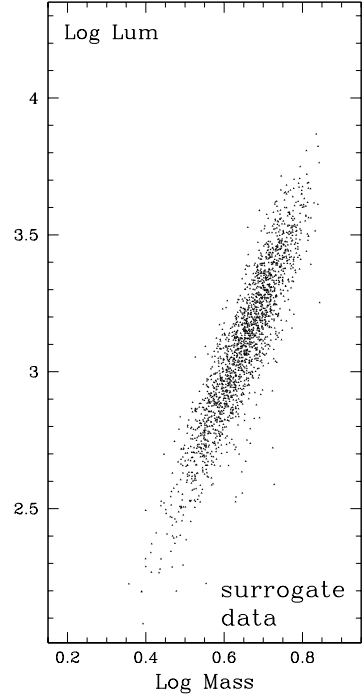}}}
 \hfill
 \vskip 0pt
 \parbox[b]{8.5cm}{
\caption[]{\small Mass--luminosity relations for surrogate stars.
  }
 \label{surrog}
 }
 \end{figure}

In order to see whether the size of error that is inherent in the
observations is responsible for the scatter in our $M$--$L$ relation
we have made the following test.  First we construct a sequence of
fundamental Cepheid models with a specific $M$--$L$ relation, $\Log
L = 0.79 + 3.56~\Log M$, and with a range of \Teff\ that spans the
corresponding instability strip.  With Eqs.~(1--6) we transform
$L$ and \Teff\ to I and V magnitudes, and maculate these data
with Gaussian noise of intensity 0.02 in the I and V magnitudes, and
with a Gaussian noise in the reddening with $\sigma_{E(B-V)}$ =
0.06.  Using these surrogate stars as input we then proceed to
compute the surrogate stellar masses the same way as we handled the
OGLE data.  Fig.~\ref{surrog} shows the resulting $M$--$L$
relation.  It is seen to exhibit the same type of scatter as the
OGLE derived $M$--$L$ relations.

This sensitivity can also be seen analytically.  For that purpose we
have made a rough fit with the help of  our models 
\begin{equation}
\Log P_0 \! =\! 11.80 \! -\! 0.595 \Log M\! +\! 0.82 \Log L\! -\! 
3.55 \Log \teff
\end{equation}
Together with Eq.~(\ref{eqs}) one then derives
\begin{eqnarray}
\delta \Log M &=& 0.51 \delta \mu - 0.51 \delta {\rm V} 
               +1.37 \delta ({\rm V-I}) \\
\delta \Log L &=& 0.4\delta \mu -0.4\delta{\rm V} 
               +1.32 \delta {\rm E(B-V)}               
\end{eqnarray}
As a check, with the noise level 0.02 in the I and V magnitudes
and 0.06 in the reddening for the preceding test one obtains
$\delta\Log M = 0.05$ and $\delta\Log L = 0.09$, in agreement with
the numerical results.

{\sl The reason for the scatter in the derived $M$--$L$ relation is thus
seen to originate in the extreme sensitivity of the mass--luminosity
relation to small photometric errors in the V and I magnitudes}.

It is very tempting to use the observational deviations in
Fig.~\ref{res} to tighten the derived $M$--$L$ data.  The question
is whether we can use the deviations parallel to the reddening line
to estimate (and correct for) the reddening and observational noise.
We find that, because of the finite width of the instability strip,
the spread in \Teff\ has an effect parallel to the reddening, so we
cannot decouple the reddening error from it.  The spread in mass
(for a given $L$) has an effect not parallel (at an angle close to
45 degrees) to the reddening, thus with a component perpendicular to
the reddening line.  Because of this projection angle the
perpendicular direction alone cannot be used to estimate the
observational errors in I or V.  In summary, it is unfortunately not
possible to use the residuals of Fig.~\ref{res} to correct for the
observational reddening errors on a star by star basis.


 \begin{figure*}
 \centerline{\resizebox{9.7cm}{!}{\includegraphics{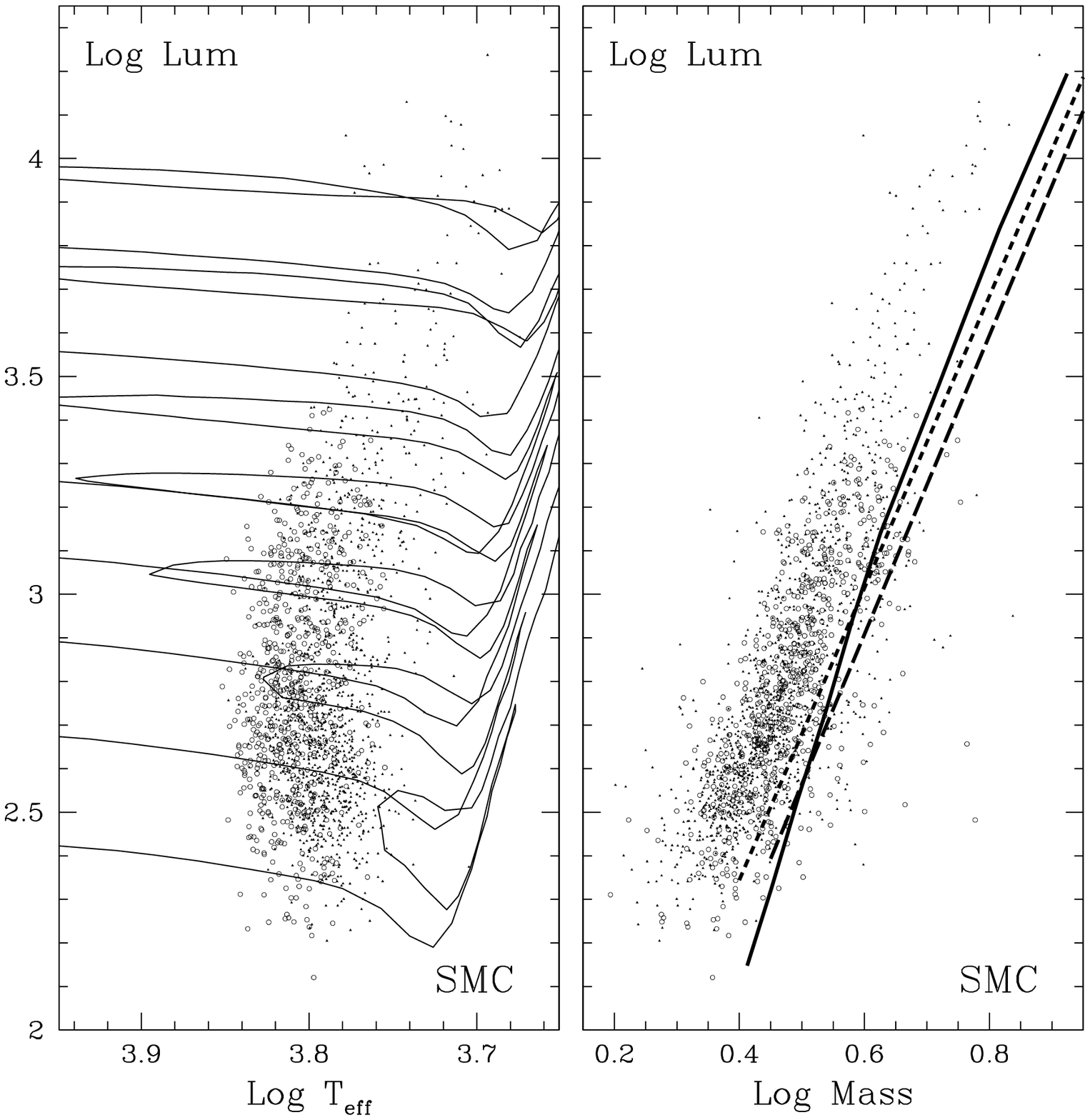}}\hskip -20pt
\resizebox{9.7cm}{!}{\includegraphics{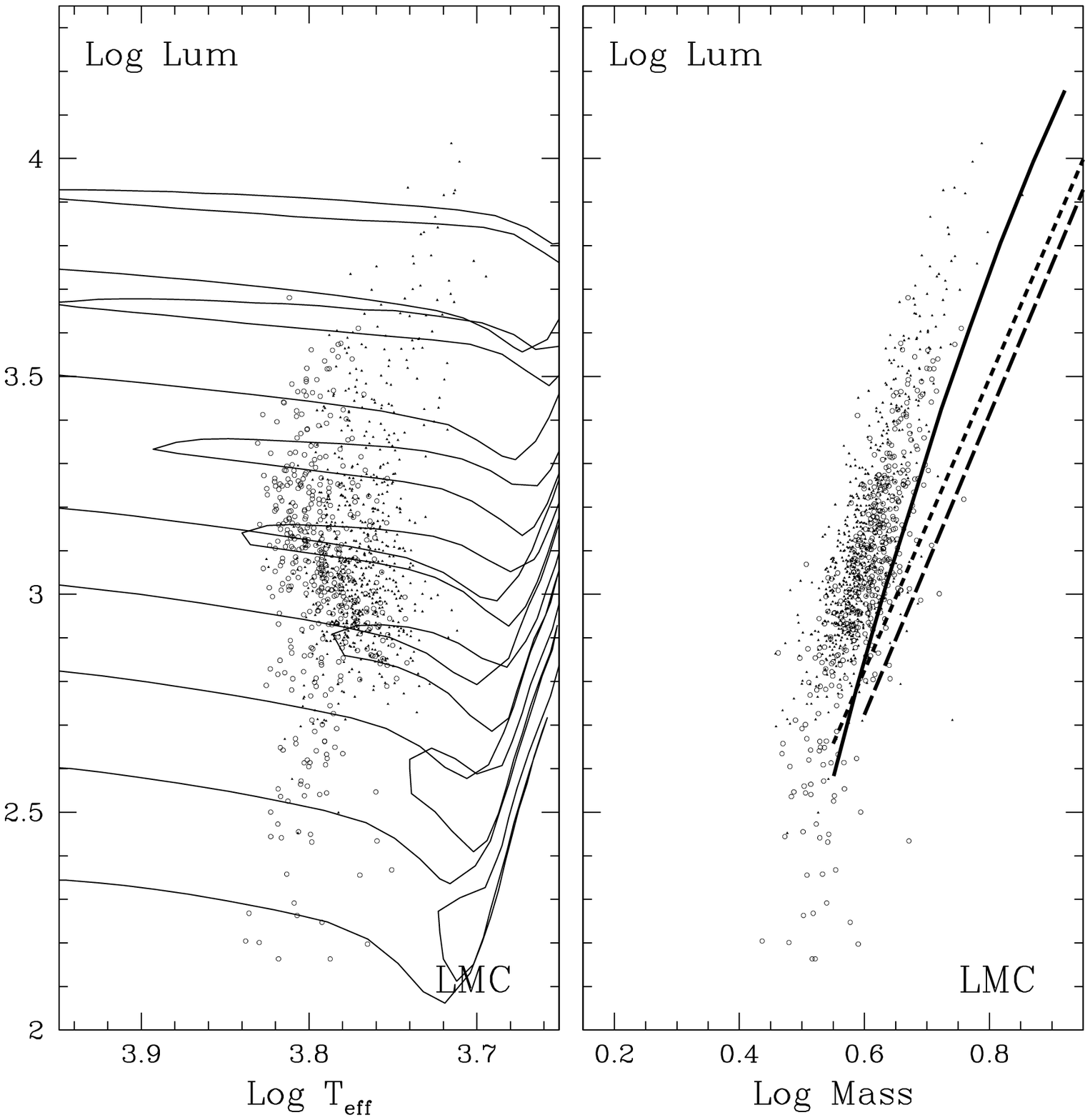}}}
 \hfill
 \vskip 0pt
 \parbox[b]{17cm}{
\caption[]{\small Theoretical HR diagram and Mass--luminosity 
relations for SMC (the two left panels) and for LMC (the two right panels).   As
in Fig.~\ref{mcab}, fundamental Cepheids are shown as solid and overtones as open
circles calculations: Theoretical HR diagram with superposed 
evolutionary tracks from Girardi \etal\ (2000).  $M$--$L$ relations from evolutionary calculations; 
solid lines: Girardi \etal\ 2000,  
dotted lines: Alibert \etal\ 1999,
dashed lines: Bono \etal\ 2000
.}
 \label{evol}
 }
 \end{figure*}

\section {Beat Cepheids}

OGLE have also published data on SMC beat Cepheids.  Because the
knowledge of a (precise) second period adds a vital piece of
information, these stars should be even more constraining than the
single-mode Cepheids for extracting an $M$--$L$ relation.  In fact
Kov\'acs (2000) has used the two observed periods and \Teff\, and
radiative linear Cepheid models to infer luminosities and thus the
distance modulus to the SMC.

 \begin{figure*}
 \centerline{\resizebox{14cm}{!}{\includegraphics{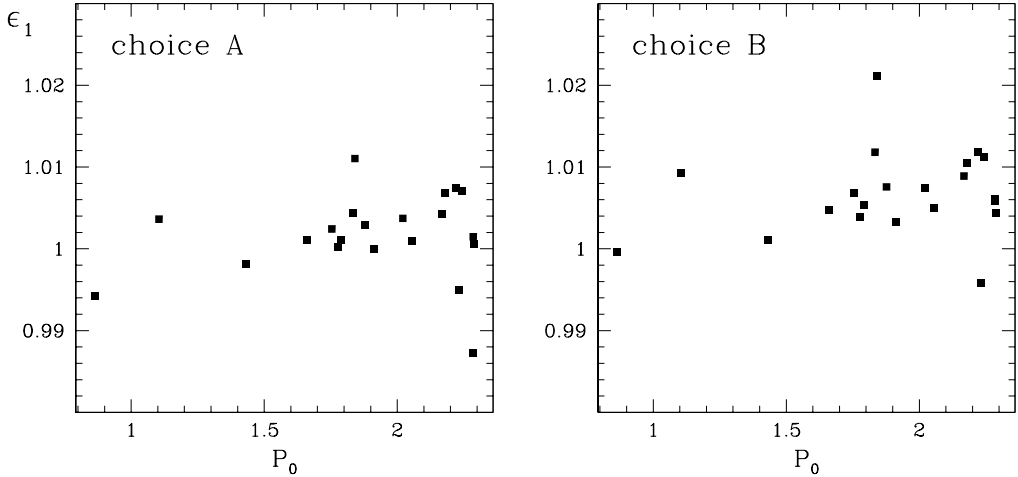}}}
 \hfill
 \vskip 0pt
 \parbox[b]{17cm}{
\caption[]{\small 
SMC F/O1 stars: 
$\epsilon_1$ = $P_{1}(calc) / P_{1}(obs)$ \vs the fundamental period for 
the SMC F/O1 stars. Left: distance modulus from
choice  (A), Right: distance modulus from choice (B).
  }
 \label{dmfigf1o}
 }
 \end{figure*}

In order to check the self consistency of the observational data and
pulsation models we can make the following test on the SMC beat
Cepheids.  We take three of the four observed quantities, \viz
\Teff, $L$, $P_k$ and $P_{k+1}$ ($k=0$ for the F/O1 and $k=1$ for
the O1/O2 beat Cepheids).  From these three parameters (ignoring
$P_{k+1}$ for the time being) we calculate the mass and then the
second period $P_{k+1}(calc)$.  Then we compare this calculated
period to the observed one ($P_{k+1}(obs)$) in Fig.~\ref{dmfigf1o} 
and Fig.~\ref{dmfig}.  On
the $\epsilon$ = $P_{k+1}(calc) / P_{k+1}(obs)$ \vs $P_k$ diagram,
with the choices (A, B, C) of distance modulus and $E(B-V)$ we observe the
following facts:

 \begin{figure*}
 \centerline{\resizebox{14cm}{!}{\includegraphics{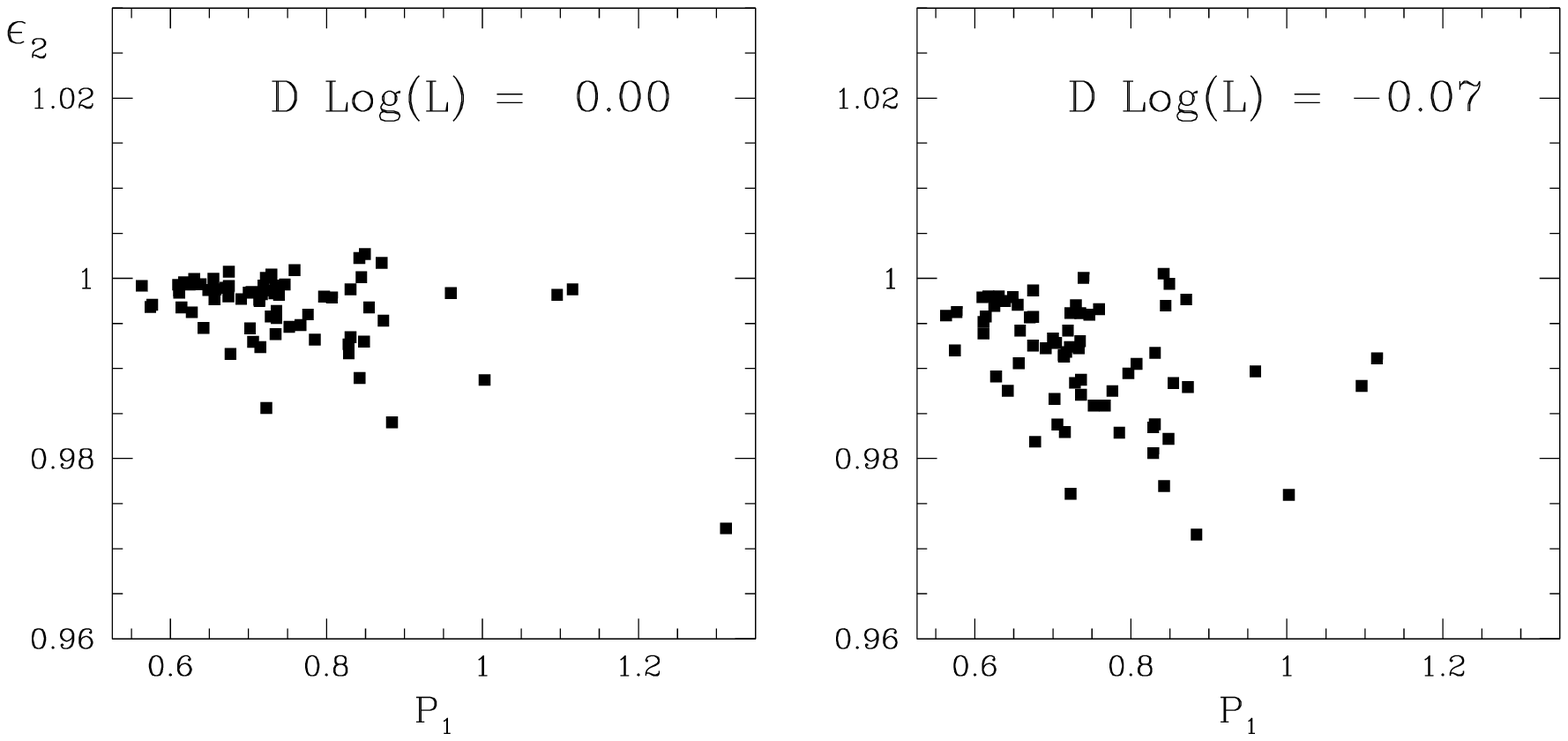}}}
 \centerline{\resizebox{14cm}{!}{\includegraphics{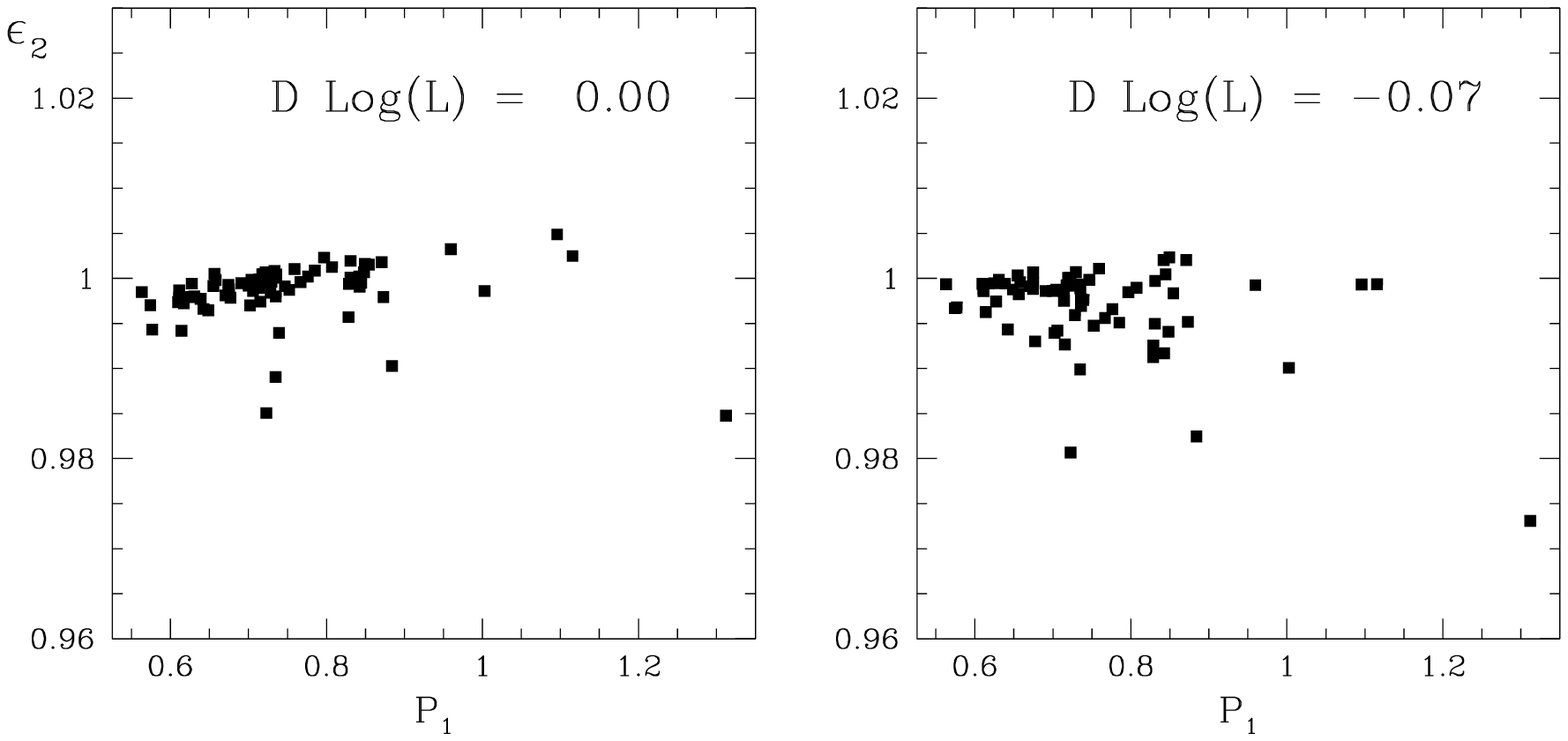}}}
 \hfill
 \vskip 0pt
 \parbox[b]{17cm}{
\caption[]{\small 
SMC O1/O2 stars:  $\epsilon_2$ = $P_{2}(calc) / P_{2}(obs)$ \vs the first
overtone period for the SMC O1/O2 stars.
The upper pannels are for the choice (A) of distance and reddening, whereas the lower pannels 
are for the choice (B).
 Left: distance modulus from choice (A) or (B), Right: with $\delta {\rm Log} L =-0.07$.
  }
 \label{dmfig}
 }
 \end{figure*}

\begin{itemize}

\item{} For the F/O1 stars, the data fall along an almost horizontal line with
little scatter.
A similar result is obtained for the different choices, but $\epsilon$ 
is closer to 1 for choice (A). To get a self-consistent
solution for these stars for choice (B), we have to increase the luminosity
by $\delta {\rm Log} (L) =
0.12$. Decreasing the metal content ($Z$) to 0.001 also shifts the
$\epsilon$ values in the right direction, but by itself it does not
solve the discrepancy.

\item{} For the O1/O2 beat Cepheids, the $\epsilon$ values scatter around
a line with a slope of $\approx 0.03$. There is only a limited range
around $P_1 \approx$ 1.0 days, where self consistent solutions exist
for the stellar parameters.

\item{} The scatter on the $\epsilon$ vs. $P_k$ plots are consistent
with the observational noise in $E(B-V)$, I and V. 
We note that with the help of
surrogate data with the same noise as described in Sect. 2.3, we
found that these error sources do not introduce systematic trends
(like the slope of $\epsilon$).

\end{itemize}

For the second set of tests we allowed systematic shifts in Log $L$.
For the O1/O2 Cepheids, the slope of $\epsilon$ strongly depends on
the adopted $\delta {\rm Log} L$. With $\delta {\rm Log} L$ =
$-0.05$ to $-0.10$, the slope is removed but the scatter of the
points is increased, and $\epsilon < 1$ for all of the
stars. Consistent solutions exist again only if the metallicity
($Z$) is decreased to 0.001.  In the case of F/O1 stars the distance
modulus has a less significant effect on the slope of
$\epsilon$. The best agreement was found with $\delta {\rm Log} L =
0.10 $ which is  opposite to the value we found for the O1/O2
Cepheids.

We have checked whether this discrepancy can be removed by allowing
a wider range of initial assumptions on the input parameters.
For our first set of tests the distance modulus was fixed, and we
allowed a wide range in reddening ($-0.1 < \Delta E(B-V) < 0.1$) as
well as various changes in the composition and metallicity mixtures
with the customized OPAL library.  All these changes in the input
data result in some vertical shifts in the $\epsilon$ \vs $P_k$
diagram, but not enough to get consistent solutions for the F/O1
stars.  The metallicity would need to be decreased to $Z=0.001$ to
get the mean value of $\epsilon$ to be 1.  We also note that there
is no significant difference in $\epsilon$ between the radiative and
convective models.

Our conclusion agrees with the work of Buchler, Koll\'ath, Beaulieu
\& Goupil (1996), but is in apparent disagreement with Kov\'acs
(2000).  The reason for this apparent disagreement is that Kov\'acs
did not construct models with the observational parameters,
but simply minimized what he called $\sigma$, \viz his measure of 
the deviation
from observed to model periods, and in fact this sigma is not zero
for many of his 'solutions'.  Furthermore in those cases where a solution
can be found, the mass is determined with a very large uncertainty
by the two period constraint, as already pointed out by Buchler
\etal\ (1996).

Moreover, although this does not directly affect the absence of
solutions, we remark that Kov\'acs adopted reddening following that of U99.  
These reddenings are $\sim 0.01$ larger than the mean reddening towards the SMC.  
This will
marginally affect his temperature scale compared to ours. However
the distance he derives is not in agreement with the distance
adopted by U99 to the SMC, but is close to ours. 

We note that the same trouble arises when we use the 3 observational
data, ($P_k,$ $P_{k+1}$, $\teff$) and compute $L$ and $M$.  For many
stars in the SMC sample there is no solution, \ie no mass and
luminosity can be found that satisfies these three observational
constraints!  The same difficulty appears when, instead, one tries
to satisfy the 3 observational constraints ($P_k, P_{k+1}, L$) to
compute a \Teff\ and $M$.

In the cases where there are solutions based on three pieces of
observational data, they are generally not compatible with the
fourth one, \ie if the periods and \Teff\ are given, the calculated
luminosity and mass are not fully acceptable.  Why there are no
satisfactory solutions for the observed beat Cepheids in the SMC
remains an unsolved puzzle that the introduction of turbulent
convection in the linear codes has not resolved.

\section{Comparison to Evolutionary Tracks}

It is of course of great interest to confront the predictions of
evolution calculations with the $M$--$L$ values which we have
extracted from the OGLE data.  Recently a number of such
calculations, all performed with the OPAL opacities, have become
available: Alibert \etal\ (1999), Girardi \etal\ (2000), Bono \etal\ 
(2000).  Alibert \etal\ have also compared their results to
observational EROS $P$--magnitude--color values and have obtained
reasonable agreement.  Of course our results, by construction,
satisfy these observational constraints.

The evolution calculations have not been compared to the SMC and LMC
OGLE-derived stellar parameters, neither in a HR diagram nor in a
$M$--$L$ diagram, both of which we now present in Fig.~\ref{evol}.
We plot the evolutionary tracks of Girardi with ($X$=0.756,
$Z$=0.004) and ($X$=0.742, $Z$=0.008), respectively, which are the
closest to our chosen compositions.  The Girardi \etal\ $M$--$L$
relations for the 2nd/3rd crossing (taken as the points of slowest
evolution near the blue edge) are shown as solid lines, those of
Alibert \etal\ as long dashes and those of Bono \etal\ with
($Y$=0.226, $Z$=0.004) and ($Y$=0.216, $Z$=0.004) as short dashes.

We note that none of the three sets of evolutionary calculations is
fully in agreement with our OGLE-derived LMC and SMC $M$-$L$ data
(nor are the older calculations), even when we adopt the most
favorable choice of distance modulus and reddening.  At fixed mass,
the evolved stellar models are not luminous enough.  The results of
Girardi \etal\ are closest to our derived $M$--$L$ relations, and
they also seem to have the right curvature (Alibert \etal\ and Bono
\etal\ used straight line $M$--$L$ fits).  Indeed, if the $M$--$L$ of
Girardi \etal\ are shifted by $\sim$ 0.35 in $\log L$ for SMC
(respectively by $\sim$ 0.25 in $\log L$ for LMC) metallicities, a
reasonable agreement obtains at low and high luminosities.  This
could be achieved, at least partially, with overshooting or an
increase thereof (Baraffe, priv. comm.).

We have not shown the $M$--$L$ relations for the faster, first
crossing to avoid cluttering the figures.  It can be seen from the
left-hand sub-figures that the luminosities are about 0.2 lower for
the same mass on these crossings.

The density of stars is definitely lower at the low luminosity end.
A natural explanation is that the low luminosity stars are first
crossers.  The Girardi \etal\ tracks for the LMC are compatible with
this interpretation, but it would be useful to do the statistics on
the basis of the evolution speed along the tracks.  However, for the
SMC, even the Girardi $M$--$L$ is much too low for the first
crossers.

Finally, the Girardi low $L$ tracks do not loop sufficiently far for
the SMC.  The problem is slightly worse for both the LMC and SMC
tracks of Alibert \etal.

The reader may wonder why both the evolutionary calculations and our
pulsational calculations claim to give agreement with the
observational data, yet they are based on substantially different
stellar masses.  To discuss the origin of this discrepancy we first
need to compare the two procedures.

In our calculations, we rely only on observed periods, colors and
magnitudes, transformed to \Teff, $L$ with Kurucz models, and from
which we construct stellar models that exactly satisfy these
observational constraints.  The computed periods are essentially
independent of any physical and numerical uncertainties, as pointed
out in \S3.2.  Furthermore, we do not rely on the stability of the
models.  The latter are quite sensitive to physical uncertainties,
especially the turbulent convective parameters.  (We note though
that, with our 'standard' parameters, the vast majority of our
models are linearly unstable.)

In order to determine the \Teff\ range of Cepheid behavior along the
evolutionary tracks, Alibert \etal, for example, performed a linear
stability analysis along these tracks and then compared the
properties of their {\it unstable} models to the observational EROS
$P$--magnitude and $P$--color data sets.  The use of a stability
analysis with the inherent uncertainties coming mostly from
turbulent convection necessarily introduces an uncertainty in
the temperature range of the Cepheid models (width of the
instability strip).

Equation (7) gives a clue to the apparent discrepancy: It is
possible to absorb rather large changes in $M$, say $\delta \Log M =
0.1$ (the largest difference between the evolutionary calculations,
or with our average $M$ -- $L$), with a tiny change in $\Log\teff$
of $\delta\Log\teff = 0.02$.  This shows that our procedure of
extracting masses directly from the observational data can therefore
impose a novel, external and stronger constraint on the evolutionary
calculations than is available from the previous comparison to
observed $P$--magnitudes and colors.

\section{Conclusions}

We have used the OGLE data base and Kurucz atmosphere models to
obtain periods, effective temperatures and luminosities for
fundamental and overtone Magellanic Cloud Cepheids.  With an assumed
average composition of ($X$=0.716, $Z$=0.010) for the LMC and of
($X$=0.726, $Z$=0.004) for the SMC, stellar masses have been computed
with the Florida pulsation code for the observationally derived $P$,
$L$ and \Teff.  The best results are obtained by adopting the 'long'
distances to the clouds.  Reddening maps, as opposed to average MC
reddenings, do not reduce the unexpectedly large scatter in
$M$--$L$, and we conclude that individual stellar reddenings would
be necessary.  However, the largest source of the scatter is in the
V and I photometric errors.  The $M$--$L$ relations for the
fundamental and for the first overtone Cepheids match closely for
each Magellanic Cloud.  Both the SMC and the LMC $\Log M$--$\Log L$
relations have a noticeable curvature.

A comparison with the predictions of the recent stellar evolution
calculations show a discrepancy in the $M$--$L$ diagrams, where the
evolutionary tracks are underluminous.  It also confirms a
discrepancy in the theoretical HR diagrams, especially for the SMC,
where the low mass tracks do not extend sufficiently blue-ward to
penetrate the instability strip.
Our $M$--$L$ relations are useful as a new type of constraint that
the evolutionary tracks have to satisfy.

\begin{acknowledgements}

This work has been supported by the National Science Foundation
(AST98-19608) and by the Hungarian OTKA (T-026031). It is based on observations coming 
from the public archive from the OGLE-II microlensing experiment. We have
profited from a fruitful correspondence with G\'eza Kov\'acs,
Isabelle Baraffe, Marie-Jo Goupil, Daniel Cordier, Giuseppe Bono.
We thank Andrejz Udalski for his comments on the paper.  Two of us
(JRB and ZK) gratefully acknowledge the hospitality of the Institut
d'Astrophysique de Paris, as a ``Professeur invit\'e de
l'Universit\'e Paris VI'' (JRB), and as a ``Poste rouge'' (ZK).

\end{acknowledgements}

\end{document}